\documentclass[12pt]{iopart}

\usepackage{iopams}  

\usepackage{graphicx}
\usepackage{epsfig}

\begin{document}

\title[Temporal Coherence of Spatially Indirect Excitons]
{Temporal Coherence of Spatially Indirect Excitons across Bose-Einstein Condensation: the Role of Free Carriers}
\author{Romain Anankine$^{1}$, Suzanne Dang$^{1}$, Mussie Beian$^{1}$, Edmond Cambril$^{2}$, Carmen  Gomez Carbonell$^{2}$, Aristide Lema\^{i}tre$^{2}$, Fran\c{c}ois Dubin$^{2}$}

\address{1 -- Sorbonne Universit\'{e} UPMC, CNRS-UMR 7588, Institut des NanoSciences de Paris, 4 Place Jussieu, F-75005 Paris, France\\
2 -- Centre for Nanoscience
and Nanotechnology -- C2N, University Paris Saclay and CNRS, Route de
Nozay, 91460 Marcoussis, France}
\ead{francois.dubin@insp.upmc.fr}
\vspace{10pt}

\begin{abstract}
We study the time coherence of the photoluminescence radiated by spatially indirect excitons confined in a 10 $\mu$m electrostatic trap. Above a critical temperature of 1 Kelvin, we show that the photoluminescence has a homogeneous spectral width of about 500 $\mu$eV which weakly varies with the exciton density. By contrast, the spectral width reduces by two-fold below the critical temperature and for experimental parameters at which excitons undergo a gray Bose-Einstein condensation. In this regime, we find evidence showing that the excitons temporal coherence is limited by their interaction with a low-concentration of residual excess charges, leading to a minimum photoluminescence spectral width of around 300 $\mu$eV.
\end{abstract}

\section{Introduction}

In semiconductor physics, Bose-Einstein condensation of excitons, i.e. Coulomb bound electron-hole pairs, has received a large attention since original theoretical predictions were formulated in the 1960's \cite{Blatt_62,Moskalenko_62,Keldysh_68}. Indeed, excitons are very attractive because their effective mass, much lighter than atoms, yields a cryogenically accessible critical temperature for the phase transition. However, the excitons electronic structure has hidden for a long time their condensation \cite{Combescot_2007}. This property is well illustrated in widely studied GaAs heterostructures  where excitons have four accessible "spin" states, two optically dark states with spins ($\pm$2) and two optically bright ones with spins ($\pm$1). Bright excitons lying at a slightly higher energy than dark ones  \cite{Blackwood_94,Mashkov_97},  Bose-Einstein condensation leads to a macroscopic occupation of lowest energy dark states so that the condensate is not detectable easily. In fact, it is only in double quantum wells that a condensate can be imaged directly \cite{Shiau_2017}. Such heterostructures allow to enforce a spatial separation between electronic carriers so that a small fraction ($\sim20\%$) of bright excitons is possibly introduced coherently in a dark condensate by exciton-exciton interactions at sufficiently high densities ($\sim$10$^{10}$ cm$^{-2}$) \cite{Shiau_2017,Combescot_2012,Combescot_ROPP}. The  condensate then becomes gray and is signalled by the weak photoluminescence it radiates.

Recently, we have reported a series of experiments combining the signatures expected for a gray condensate of indirect excitons \cite{Lozovik} in GaAs double quantum wells (DQWs). These signatures include first a darkening of the photoluminescence below a critical temperature of about 1K and for a narrow range of densities only ($\sim$ 2-3 10$^{10}$ cm$^{-2}$) \cite{Beian_2015}. This behaviour marks a quantum statistical occupation of dark excitons because the energy splitting between bright and dark states amounts to only a few $\mu$eV in DQWs \cite{Blackwood_94,Mashkov_97}, i.e. at least 10 times less than the thermal energy at 1K. We have then confirmed the gray nature of the condensation, i.e. the coherent coupling between dark and bright excitons. For that purpose we used spatial interferometry and showed that the weak photoluminescence radiated in the regime of quantum darkening exhibits macroscopic (quantum) spatial coherence \cite{Anankine_2016}, below the same critical temperature of about 1 K and for the same narrow density range.

Here, we scrutinise another fingerprint expected for exciton condensation, namely a steep increase of temporal coherence across the condensation threshold \cite{Combescot_ROPP}. This behaviour is understood by noting that in the condensed phase optically bright excitons merely occupy \textbf{k}$\sim$0 states, directly coupled to photons. At sub-Kelvin temperatures, the excitons coherence time is then limited by interactions between condensed and uncondensed (\textbf{k}$\neq$0) excitons, as well as interactions between excitons and excess carriers which are strong in GaAs heterostructures \cite{Honold_89}. Indeed, we note that exciton-phonon interactions are negligible in the sub-Kelvin regime \cite{Singh_2000}. Bose-Einstein condensation then has to increase the excitons coherence time because interactions between excitons of the condensate can not contribute to collisional broadening due to energy and momentum conservation. In the following, for an experimentally measured condensate fraction of about 80$\%$ \cite{Anankine_2016} we show that the photoluminescence temporal coherence is increased by two-fold at threshold, the scattering rate of \textbf{k}$\sim$0 bright excitons being then reduced by the same fraction. Our studies indicate that this amplitude is limited by the amount of unintentional doping during the epitaxial growth of our heterostructure, which provides a reservoir of free carriers dephasing lowest energy excitons. 
 
\section{Experimental procedure}

We report time and spatially resolved interferometry to probe the temporal coherence of indirect excitons confined in a 10 $\mu$m trap. We study the same device as in Refs. \cite{Beian_2015,Anankine_2016} in the regime where we have shown signatures of excitonic condensation. In this heterostructure, indirect excitons (IXs) are made by the Coulomb attraction between electrons and holes confined in two 8 nm GaAs quantum wells  separated by a 4 nm AlGaAs barrier. The trap is engineered  by using two metallic electrodes, a central 10 $\mu$m wide disk surrounded by an outer guard, both deposited at the surface of a field-effect device embedding the double quantum well. We apply the same potential on both electrodes (-4.7 V), so that the 200 nm gap between them creates a rectification of potential leading to an electrostatic barrier, i.e. a hollow-trap \cite{Anankine_2016}. Moreover, note that IXs exhibit a long lifetime ($\sim 100$ ns)  in our studies and reach thermal equilibrium \cite{Ivanov_2004} in a mostly electrically neutral environment about 100 ns after optical injection, when a transient photocurrent ($\lesssim$100 pA D.C.) is evacuated. Indeed, we inject electronic carriers optically, using a 100 ns long laser excitation, repeated at a rate of 1.5 MHz and tuned on resonance with the direct exciton absorption of the two quantum wells (Fig.1). IXs are thus obtained once electrons and holes have tunnelled towards their minimum energy states, located in each quantum well. Let us then note that the laser beam covers most of the trap area.

\begin{center}
\begin{figure}[h!]\label{fig2}
\centerline{\includegraphics[width=.6\textwidth]{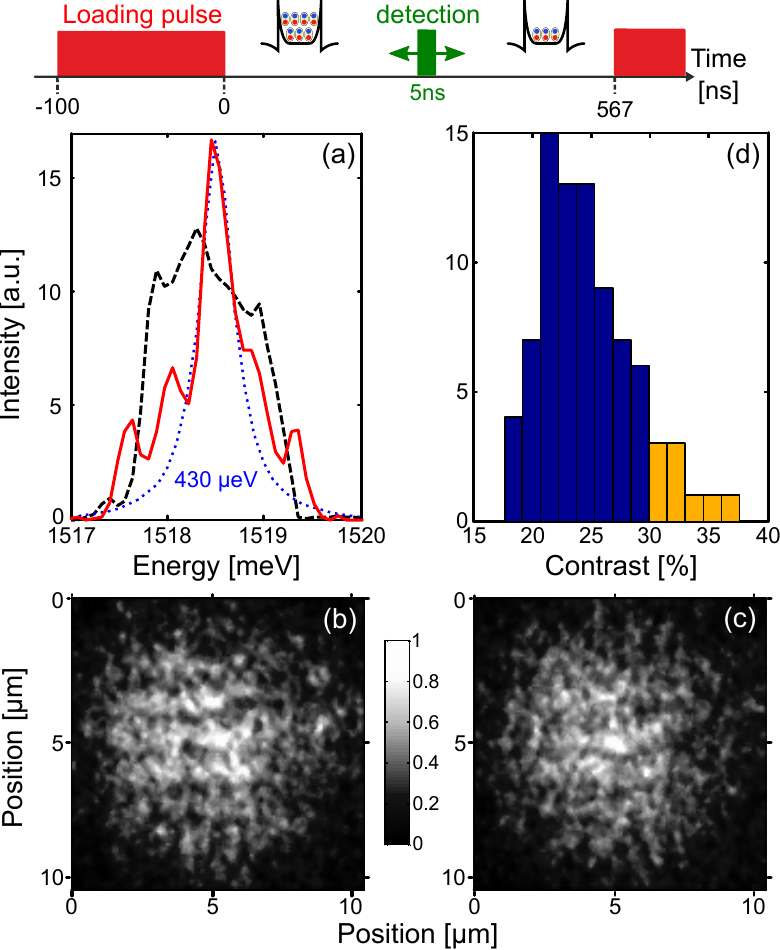}}
\caption{Our experimental sequence starts with a 100 ns laser excitation loading indirect excitons in the trap. The reemitted photoluminescence is then analysed in a 5 ns long time interval, at a given delay to the end of the loading laser pulse so that the density in the trap is varied. The sequence is repeated at 1.5 MHz. (a) Spectrally resolved PL emitted across the center of the trap at $T_\mathrm{b}=330$ mK. Measurements are realised during a 20-second long acquisition and for an exciton density $n_X\sim$ 2.5 10$^{10}$ cm$^{-2}$. The sharp line is fitted with a Lorentzian shown by the dotted line (FWHM = 430 $\mu$eV), whereas the dashed line is anomalously broad ($\sim$1 meV). High contrast (b) and lower contrast (c) interference patterns measured for the same experimental conditions as in (a), but during a 12-second long acquisition, the interferometer path length difference being set to $\delta t=1.8$ ps. (d) Distribution of interference contrasts for a sample of 100 acquisitions measured in the same experimental conditions as in (b-c). The best 10\% of these realisations is highlighted in yellow while a minimum contrast ($\sim$15\%) is set by the signal to noise ratio at the detection.}
\end{figure}  
\end{center}

To study the temporal coherence of bright IXs confined in the trap, we magnified the real image of the photoluminescence (PL) and sent it to a Mach-Zehnder interferometer. The PL is divided equally between a fixed and a mobile arm, and the two beams are superposed at the output of the interferometer, i.e. without any spatial displacement. Horizontal interference fringes are induced by a vertical tilt angle between the two outputs, of about 25$^{\circ}$. To evaluate the temporal coherence of the PL, we vary the longitudinal path length difference, i.e. we translate one arm by $\delta L$. The two outputs are then shifted temporally by $\delta t=2\delta L/c$, where $c$ is the speed of light in air, the path length difference being actively stabilised with $\sim$20 nm precision. At the output, the contrast of the interference pattern measures directly the coherence time $\tau_c$ of optically bright excitons. Indeed, the interference visibility is given by the modulus of the first-order correlation function $|g^{(1)}(\delta t)|\propto|\langle\psi^*_0(\textbf{r},t)\psi_0(r,t+\delta t)\rangle_t|$. Here, $\psi_0(\textbf{r},t)$ is the photoluminescence field which reflects the wave function of bright excitons with an in-plane momentum \textbf{k}$\sim$0 since the exciton-photon coupling is linear close to resonance \cite{Combescot_book}. Moreover $\langle..\rangle_t$ denotes the time averaging and $\textbf{r}=(x,y)$ reads the coordinate in the plane of the electrostatic trap. For a homogeneously broadened gas, the  $g^{(1)}$-function is exponentially decaying with $\delta t$, at a rate  inversely proportional to the \textbf{k}=0 excitons coherence time $\tau_c$. In a thermal regime above a few Kelvin, $\tau_c$ is controlled by exciton-phonon interactions  \cite{Singh_2000};  however, below a few-Kelvin, we expect a dramatic change because these interactions become inefficient.

\section{Spectral fluctuations and excitons coherence time}

Unlike studies performed with atomic gases \cite{Bloch_2008}, our experiments require an accumulation of several millions of single-shot realisations, during around 10 to 20 seconds, each realisation starting by a loading laser pulse after which the PL is interferometrically analysed (Fig.1). This approach is inevitable due to the weak photoluminescence signal radiated from the trap, but it constitutes a severe limitation. Indeed, we observe large spectral fluctuations, with a characteristic timescale of about 10 sec, as shown in Fig. 1 at a bath temperature T$_\mathrm{b}$=330 mK and for a trapped exciton density n$_\mathrm{X}$ about 2.5$\cdot$10$^{10}$ cm$^{-2}$, obtained for a delay of 170 ns after the loading laser pulse. Particularly, Fig.1.a displays two spectra measured successively; a first  spectrum consists mostly of a central narrow line, with a width limited by the spectrometer resolution. By contrast, another spectrum obtained for the same experimental settings can be about 1 meV broad with no recognisable shape (dashed line in Fig.1.a). 

We attribute the spectral diffusion evidenced in Fig.1.a as the result of the large electric dipole moment ($\sim$ 500 Debye) characterising indirect excitons and which makes the energy of their photoluminescence very sensitive to the electrostatic environment. On the one hand this property is useful since it allows us to extract the density of  excitons in the trap n$_X$, indirect excitons experiencing repulsive dipolar interactions that yield a blue shift of the photoluminescence energy scaling as $u_0$n$_X$ \cite{zimmermann2008bose,Ivanov_2010}, where $u_0\sim$0.5 meV/10$^{10}$ cm$^{-2}$. On the other hand, indirect excitons interact strongly with free carriers, due to either the laser induced photo-current or to the level of residual doping during the growth of our heterostructure. In both cases, the exciton-free carrier  interactions provide a direct source for spectral diffusion because the local concentration of excess carriers can not be controlled and \textit{a priori} varies randomly during our measurements. To minimise this effect, we typically analyse the PL long after extinction of the laser excitation, i.e. when the photo-current is damped. Nevertheless, even a density of excess carriers small compared to n$_X$ can yield a significant electrostatic noise, because in GaAs quantum wells the interaction strength between excitons and free carriers is about 10 times larger than the one between excitons \cite{Honold_89}.

Spectral fluctuations are observed directly in the temporal interference signal, with actually better dynamics since the acquisition time is reduced from about 20 to 10 seconds once the PL is not filtered spectrally. For fixed experimental conditions at T$_\mathrm{b}$=330 mK,  Fig.1.b shows that either a clear interference pattern can be observed across the center of the trap, or a rather blurred pattern (Fig.1.c). For these experiments where the path length difference is set such that $\delta_t$=1.8 ps, in the former case the spectrum is expected narrowband ($\sim$300 $\mu$eV) whereas it is wide in the latter case ($\sim$700 $\mu$eV). The spectral diffusion is further evidenced in Fig. 1.d where we display the interference visibility measured at T$_\mathrm{b}$=330 mK for a statistical ensemble of 100 experiments performed all under the same conditions. We note that the contrast exhibits an asymmetric dispersion, from 17 to 37 $\%$, marked by a tail at high contrasts (30 to 37 $\%$) and a sharp growth at low contrasts (from 17 to 22 $\%$). While we can not quantify the origin of such asymmetry, it indicates that the spectral diffusion is complex and not simply due to a gaussian noise process. Based on the calibration of our interferometer, we nevertheless deduce that the photoluminescence exhibits a spectral width systematically lower than 700 $\mu$eV and that for the most regular electrostatic environment, e.g. for a lowest concentration of excess free carriers, the spectral width can become as narrow as 300 $\mu$eV. 

\section{Excitons temporal coherence}

We now study the variation of the photoluminescence coherence time as a function of the exciton density n$_X$ and bath temperature T$_\mathrm{b}$. To limit the influence of spectral fluctuations over our conclusions, we restrict our analysis to the realisations with highest visibility for each experimental settings. Precisely, we average the contrasts measured for the best 10$\%$ of experimental realisations, post-selected within a statistical ensemble of a minimum of 50 measurements for each experimental conditions (see for instance the orange coloured region in Fig.1.d). Thus, we aim at minimising the amount of measurements where inhomogeneous broadening plays a significant role. Instead, we focus onto experiments where indirect excitons are confined in the most regular trapping potential, combined to a minimum amount of excess  charges interacting with them. Post-selecting these optimum conditions is actually necessary to study the excitons gray condensation, since we have shown in Ref.\cite{Anankine_2016} that a gray condensate is only obtained for such  optimised experimental settings and can not form otherwise.

Let us first discuss the dilute regime,  i.e. an exciton concentration $n_X$ of about 2.5$\cdot$10$^{10}$ cm$^{-2}$ obtained for a 170 ns delay to the extinction of the loading laser pulse. Figure 2 quantifies the variation of the interference contrast along the center of the trap, as a function of the time delay $\delta t$ introduced between the two arms of the interferometer. We note that the interference visibility follows a mono-exponential decay $\mathrm{e}^{-\vert\delta t\vert/\tau_c}$, at every bath temperature T$_\mathrm{b}$ below 2.36 K. According to the Wiener-Kintchine theorem \cite{Born_2000}, the photoluminescence spectrum then consists of a single Lorentzian. This shows that we study a homogeneously broadened gas which constitutes a major improvement compared to previous works \cite{High_2009,Holleitner_2013,Rapaport_2016,Repp_2014,Stern_2014,Timofeev_2016}. To the best of our knowledge, such a single Lorentzian PL profile had never been observed for a gas of indirect excitons. For the measurements shown in Figure 2, the photoluminescence spectral width $\Gamma$=$2\hbar/\tau_c$ reaches then 550 $\mu$eV at T$_\mathrm{b}$=2.4 K. This value is nevertheless $\sim$300 $\mu$eV greater than the theoretically expected linewidth due to exciton-exciton interactions \cite{zimmermann2008bose}, indicating that these latter are not the only mechanism governing the homogeneous broadening.

\begin{center}
\begin{figure}[h!]\label{fig3}
\centerline{\includegraphics[width=.6\textwidth]{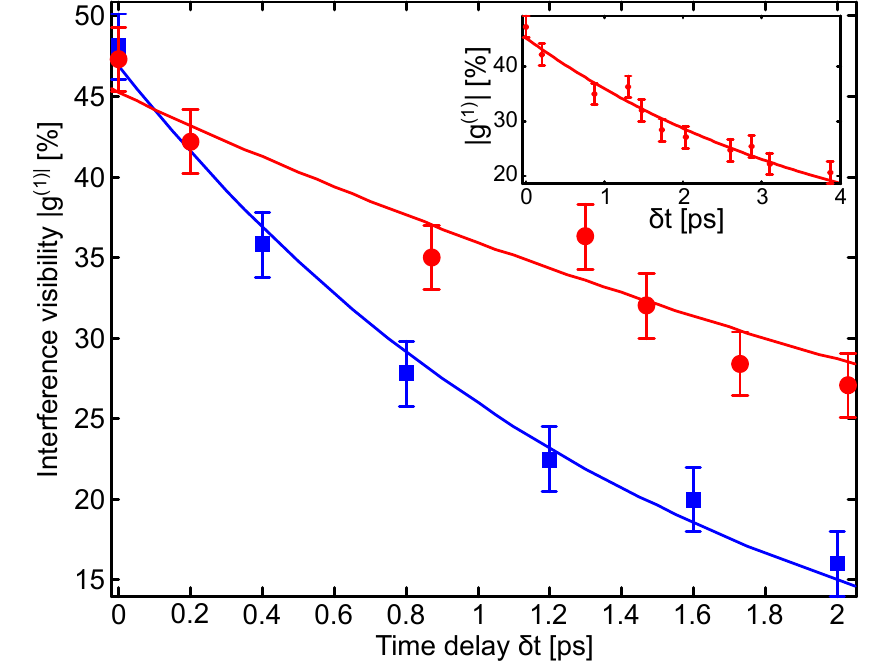}}
\caption{Interference visibility measured along the center of the trap as a function of the time delay introduced in the interferometer. Measurements taken between $T_\mathrm{b}=330$ mK (red circles) and 2.36 K (blue squares) all follow single exponential decays (solid lines). The insert displays the data at $T_\mathrm{b}=330$ mK with an extended abscissa, fitted by a single exponential with a time decay of 4 ps. }
\end{figure}
\end{center}

Varying the delay to the loading laser pulse we study directly the photoluminescence temporal coherence as a function of the exciton density. Indeed, the latter decreases monotonously in the trap due to radiative recombination, and let us stress that in our studies the total density of excitons n$_\mathrm{X}$, including both bright and dark states, is the same at a given delay for every bath temperature below 3K \cite{Beian_2015}. For a fixed delay $\delta t$=1.8 ps, Figure 3 shows that three regimes emerge while the exciton density is varied: (i) In the most dilute regime, i.e. for longest delays to the end of the loading laser pulse, the interference contrast does not vary with T$_\mathrm{b}$. This behaviour indicates that the gas is localized, as expected since in this range of densities the blueshift of the photoluminescence, i.e. the excitons mean interaction energy, does not exceed the electrostatic disorder in the trap ($\sim$500 $\mu$eV). At higher densities (regime (ii)) the excitons mean-field energy compensates the strength of potential disorder so that dipolar repulsions delocalise the exciton gas. Then, the interference contrast varies strongly with T$_\mathrm{b}$: while the visibility keeps a value similar to regime (i) for T$_\mathrm{b}\geq$1.2 K, it increases steeply at  T$_\mathrm{b}$=330 mK, from about 24 to 30 $\%$ when $n_X$  $\gtrsim$2$\cdot$10$^{10}$ cm$^{-2}$. Increasing further the density, $n_X$  $\gtrsim$  3.5$\cdot$10$^{10}$ cm$^{-2}$ in regime (iii), leads to a breakdown of this behaviour and the interference contrast depends again weakly on the bath temperature.

\begin{center}
\begin{figure}[h!]\label{fig4}
\centerline{\includegraphics[width=.75\textwidth]{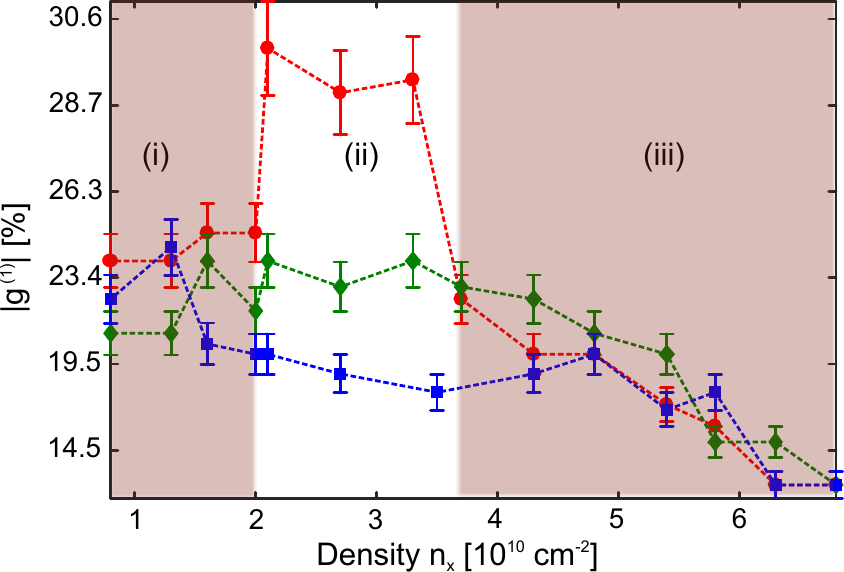}}
\caption{(a) Interference contrast $|g^{(1)}|$ measured along the centre of the trap as a function of the exciton density, $n_X$.
Measurements at $T_\mathrm{b}$=330 mK, 1.2 and 2.4 K, are displayed in red, green and blue, respectively. Regions (i) to (iii) are discussed in the main text.}
\end{figure}
\end{center}

From the visibility measured for $\delta t$=1.8 ps, we can deduce the spectral width $\Gamma=2\hbar/\tau_c$ in regime (ii) where we have shown in Fig.2 that the photoluminescence is homogeneously broadened.  At $T_\mathrm{b}$=1.2 K, Fig.3 shows that the homogeneous spectral width is about 500 $\mu$eV at 1.2K and varies weakly with the exciton density $n_X$. At $T_\mathrm{b}$=2.4 K , $\Gamma$ is slightly larger ($\sim$600 $\mu$eV), possibly due to the increased role of acoustic phonons, while it also varies weakly with $n_X$. Overall, these variations with n$_\mathrm{X}$ are theoretically expected \cite{zimmermann2008bose,Schindler_2008}, since interactions between indirect excitons vary $\Gamma$ by less than 100 $\mu$eV in the density range explored here, i.e. by about our instrumental precision. However, the  magnitude of $\Gamma$ lies about 300 $\mu$eV above its theoretical expectation. We attribute this discrepancy as a manifestation of the interaction between indirect excitons and residual excess charges, although we minimise the concentration of these latter in regime (ii) since the delay to the laser excitation exceeds 120 ns. 

\section{Excess free carriers and collisional broadening}

To study the role of excess carriers in the double quantum well, let us first estimate the fraction of free carriers induced by the laser excitation. For that purpose, we consider the dynamics of the photocurrent, which average $\bar{I}$ amounts to 100 pA, as measured in a 20 ms time interval with the source-measure device used to polarise the exciton trap.  Moreover, we have previously shown that the photocurrent has a transient time of the order of 70 ns \cite{Beian_2015}. In this case we deduce that $\bar{I}$=$f I_{m}T_{p}$ where $I_{m}$ is the maximum current generated at the termination of the laser excitation, while $f$ and  $T_{p}$ denote the 1.5MHz repetition rate of our sequence and  the length of the loading pulse, respectively. This expression for $\bar{I}$ directly converts into $n_m$, the maximum density of excess carriers induced at the termination of the laser excitation, since $n_m$=$I_{m}$/($e\pi f r^2$), where $e$ is the electron charge. For the radius $r$=5$\mu$m of our trap electrode, we conclude that $n_m$= 3.5 10$^9$ cm$^{-2}$ so that the density of photo-induced  excess carriers reduces exponentially to a few 10$^8$ cm$^{-2}$ in regime (ii) of Fig.3. On the other hand, excess carriers also result from the level of unintentional doping during the growth of our heterostructure. Indeed, the concentration of impurities for the molecular beam epitaxy of our sample is at most 10$^{15}$ cm$^{-3}$, yielding a density of free carriers of about 10$^9$ cm$^{-2}$ for two 8 nm wide quantum wells. This makes the level of unintentional doping the dominant source for free carriers long after the loading laser pulse.

To estimate the impact of excess carriers onto the photoluminescence spectral width, let us first compare the strength of collisional broadening due to exciton-exciton interactions, in single and double quantum wells. In the former case, experiments by Honold et al. \cite{Honold_89} have shown that the photoluminescence spectral width is increased by about 100 $\mu$eV when the exciton density grows from 10$^{10}$ to 2$\cdot$10$^{10}$ cm$^{-2}$. This collisional strength is of the same order as the calculations made by Zimmermann for indirect excitons in double quantum wells \cite{zimmermann2008bose}, neither of which contradict our observations where collisional broadening between indirect exitons is bound to our $\sim$100 $\mu$eV resolution (regime (ii) in Fig.3). This reasonable agreement suggests that in both geometries exciton-exciton interactions have a comparable strength. Let us then extend to double quantum wells the strength of interactions between excitons and excess carriers measured in single quantum wells, which is 8 times larger than interactions between excitons \cite{Honold_89}. To account for the 300 $\mu$eV mismatch between the homogeneous linewidth predicted in Ref.\cite{zimmermann2008bose} and our measurements at T$_\mathrm{b}\gtrsim$1.2K in the regime (ii), we thus deduce that a free carrier density of about a few 10$^9$ cm$^{-2}$ is sufficient. Remarkably, this concentration is of the order of the level of unintentional doping thus suggesting that this limitation is responsible for about 300$\mu$eV of the measured photoluminescence spectral width. Finally, let us note that by considering $n_m$ and the level of unintentional doping, together with the strength of interactions between excitons and excess carriers measured by Honold et al. \cite{Honold_89}, we deduce that the photoluminescence spectral width  at the termination of the laser excitation is mostly governed by excitons-free carriers interactions and amounts to about  of 1 meV. This value is in good agreement with our observations \cite{Beian_2015}, thus supporting further the previous assumptions. In fact, we attribute the rapid and temperature independent increase of $\Gamma$ in regime (iii) to excitons-free carriers interactions. Indeed, the latter regime corresponds to delays ranging from 120 to 20 ns after extinction of the loading laser pulse. While the exciton density is increased by analyzing shorter delays, the concentration of excess carriers is also increased since the transient photo-current is not fully suppressed and therefore $\Gamma$ increases in a temperature independent fashion governed by excitons-free carriers scatterings. 

\section{Threshold increase of excitonic coherence}

The most striking feature of Fig.3 is certainly the threshold increase of the interference contrast at $T_\mathrm{b}$=330 mK and $n_X$$\sim$2$\cdot$10$^{10}$ cm$^{2}$, so that $\Gamma\sim$ 300 $\mu$eV and remains rather constant until $n_X$$\sim$ 3.5$\cdot$10$^{10}$ cm$^{-2}$. Remarkably, this density range coincides with the one where we have demonstrated a gray Bose-Einstein condensate for identical experimental conditions \cite{Anankine_2016}. It is then natural to attribute the observed threshold as the result of the exciton condensation. Furthermore, $\Gamma\sim$300 $\mu$eV matches the contribution of interactions between excitons and excess carriers to the homogeneous broadening, thus suggesting that $\tau_c$ is governed by scatterings with free carriers interactions in this condensed regime. Finding this limitation is somehow not very surprising since longer coherence times are expected from exciton-exciton interactions alone \cite{Combescot_ROPP}.  Finally, at $T_\mathrm{b}$=330 mK Fig.3 shows that  $\tau_c$ abruptly decreases to its value at 1.2 K at the onset of regime (iii). This indicates that the gray condensation is accessible in a density range partially limited by the increased concentration of excess carriers at lower delays to the loading laser pulse. Nevertheless, we do not exclude that other processes becomes significant at $n_X$$\sim$ 3.5$\cdot$10$^{10}$ cm$^{-2}$, exciton ionisation being nevertheless excluded. 

\begin{center}
\begin{figure}[h!]\label{fig4}
\centerline{\includegraphics[width=.75\textwidth]{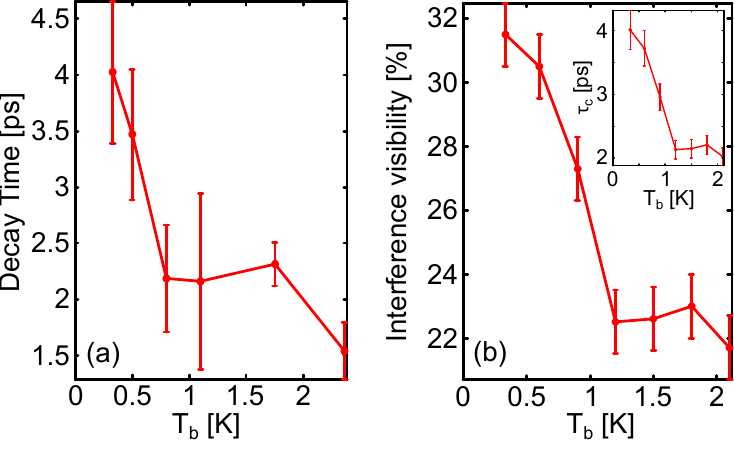}}
\caption{(a) Coherence time of the photoluminescence $\tau_c$ as a function of $T_\mathrm{b}$, deduced by fitting the exponential decay of $|g^{(1)}|$ at every bath temperature. (b) Interference visibility measured at the center of the trap for $\delta t=1.8$ ps. The inset displays the value of $\tau_c$ deduced from these measurements. All experiments have been carried out for an exciton density $n_X$$\sim$2.5 10$^{10}$ cm$^{-2}$.}
\end{figure}
\end{center}

To relate further the threshold increase of $\tau_c$ at T$_\mathrm{b}$=330 mK to the gray condensation, we measured $|g^{(1)}|$($\delta t$) as a function of $T_\mathrm{b}$ and for $n_X$$\sim$2.5$\cdot$10$^{10}$ cm$^{-2}$, i.e. at the center of regime (ii). For each bath temperature $\tau_c$ was first extracted from the exponential decay of the interference visibility with $\delta t$. As shown in Figure 4.a, a rapid increase is revealed below a bath temperature of about 800 mK, $\tau_c$ passing from about 2 ps at 1K to 4 ps at 330 mK. To measure more precisely the increase of $\tau_c$, we performed an additional experiment where the interference contrast was evaluated for $\delta t=1.8$ ps only. Thus, the stability of our measurements was largely increased since the experiments could be performed during a single day, compared to 3 days for the data displayed in Fig.4.a. As shown in Fig. 4.b we then recovered that two regimes emerge: For  $T_\mathrm{b}\geq 1.2$ K, $|g^{(1)}|$ varies weakly, it is about 22$\%$ corresponding to $\tau_c\sim 2$ ps. By contrast, below $T_\mathrm{c}\sim 1.2$ K, the interference contrast increases abruptly to 32$\%$ at $T_\mathrm{b}=330$ mK, manifesting that $\tau_c$ is doubled at sub-Kelvin bath temperatures. Let us then stress that this value of T$_\mathrm{c}$ reproduces quantitatively the critical temperature measured for the appearance of quantum spatial coherence for the same exciton density in the trap \cite{Anankine_2016}. This strongly supports our interpretation that the increase of excitonic temporal coherence is due to their quantum condensation in the trap.

\section{Conclusion}

We have shown that the gray condensation of indirect excitons is associated to a threshold increase of the photoluminescence temporal coherence. Our experiments reveal that the condensate has a coherence time limited to around 4 ps, due to interactions between indirect excitons and excess carriers.  Their concentration is very low, about 10$^{9}$ cm$^{-2}$, of the order of the level of unintentional doping during the epitaxial growth of our heterostructure. Reducing this level to a minimum value of a few 10$^{8}$ cm$^{-2}$ would lead to a maximum time coherence of a few tens of ps for gray condensates, while quantum condensation is excluded for densities of free carriers exceeding $\sim$ 5 10$^{9}$ cm$^{-2}$ \cite{Berman}. In every case, we note that the coherence time of a gray condensate remains one to two orders of magnitude smaller than for quantum fluids of polaritons \cite{LeSiDang}. We attribute this difference as the manifestation of the matter-like nature of exciton condensates which are then very sensitive to their electrostatic environment.

\textbf{Acknowledgements:} The authors are grateful to Kamel Merghem for technical support. Our experiments have been financially supported by the projects XBEC (EU-FP7-CIG), by OBELIX from the french Agency for Research (ANR-15-CE30-0020).\\


\begin{thebibliography}{99}

\bibitem{Blatt_62} J. M. Blatt et al., Phys. Rev. \textbf{126}, 1691 (1962)

\bibitem{Moskalenko_62} S.A. Moskalenko, Fiz. Tverd. Tela (Leningrad) \textbf{4}, 276 (1962)

\bibitem{Keldysh_68} L.V. Keldysh and A.N. Kozlov, Sov. Phys. JETP \textbf{27}, 521 (1968)

\bibitem{Combescot_2007} M. Combescot, O. Betbeder-Matibet, R. Combescot,
Phys. Rev. Lett. \textbf{99}, 176403 (2007)

\bibitem{Blackwood_94} E. Blackwood, M.J. Snelling, R.T. Harley, S.R. Andrews and C.T.B. Foxon, Phys. Rev. B \textbf{94}, 14246 (1994)

\bibitem{Mashkov_97} I. V. Mashkov et al. Phys. Rev. B \textbf{55}, 13761 (1997)

\bibitem{Shiau_2017} S. Shiau al., EuroPhys. Lett. \textbf{118}, 47007 (2017)

\bibitem{Combescot_2012} R. Combescot and M. Combescot, Phys. Rev. Lett. \textbf{109}, 026401 (2012)

\bibitem{Combescot_ROPP} M. Combescot, R. Combescot, F. Dubin, Rep. Prog. Phys. \textbf{80}, 066501 (2017)

\bibitem{Lozovik} Y.E. Lozovik, V.I. Yudson, Solid State Com. \textbf{19}, 391 (1976)

\bibitem{Beian_2015} M. Beian et al., EuroPhys. Lett. \textbf{119}, 37004 (2017)

\bibitem{Anankine_2016} R. Anankine et al., Phys. Rev. Lett. \textbf{118}, 127402 (2017)

\bibitem{Combescot_book} M. Combescot and S.Y. Shiau "Excitons and Cooper Pairs: Two Composite Bosons in Many-Body Physics" (Oxford. Univ. Press, 2016)

\bibitem{Singh_2000} I. K. Oh and Jai Singh, Jour. of Lum. \textbf{85}, 233 (2000)

\bibitem{Bloch_2008} I. Bloch, J. Dalibard, and W. Zwerger, Rev. Mod. Phys. \textbf{80}, 885 (2008)


\bibitem{Honold_89} A. Honold et al., Phys. Rev. B \textbf{40}, 6442 (1989)

\bibitem{Ivanov_2004} A. Ivanov, J. Phys.: Condens. Matter \textbf{16}, S3629 (2004)

\bibitem{zimmermann2008bose} R. Zimmermann, "Bose-Einstein condensation of excitons: Promise and disappointment" in {\it "Problems of Condensed Matter Physics"} (Oxford University Press, 2008, p. 285)

\bibitem{Ivanov_2010} A. L. Ivanov, E. A. Muljarov, L. Mouchliadis, and R. Zimmermann, Phys. Rev. Lett. \textbf{104}, 179701 (2010)


\bibitem{Born_2000} M. Born and E. Wolf, \textit{"Principles of optics: electromagnetic theory of propagation, interference and diffraction of light"}, (CUP Archive, 2000)

\bibitem{High_2009} A. A. High et al., Phys. Rev. Lett. \textbf{103}, 087403 (2009)

\bibitem{Holleitner_2013} G. Schinner et al., Phys. Rev. Lett. \textbf{110}, 127403 (2013)

\bibitem{Rapaport_2016} Y. Shilo et al., Nat. Comm. \textbf{4}, 2335 (2013)

\bibitem{Repp_2014} J. Repp et al., App. Phys. Lett. \textbf{105}, 241101 (2014)

\bibitem{Stern_2014} M. Stern, V. Umansky and I. Bar-Joseph,  Science \textbf{343}, 55 (2014) 

\bibitem{Timofeev_2016} A. V. Gorbunov and V. B. Timofeev, Low Temp. Phys. \textbf{42}, 340 (2016)


\bibitem{Schindler_2008} C. Schindler and R. Zimmermann, Phys. Rev. B \textbf{78},
045313 (2008)

\bibitem{Berman} O. Berman et al., Solid. Stat. Comm. \textbf{150}, 832 (2010)


\bibitem{LeSiDang} A. P. D. Love et al., Phys. Rev. Lett. \textbf{101}, 067404 (2008)





\end{thebibliography}
\end{document}